\documentclass[12pt]{article}
\usepackage{graphicx}
\usepackage{lscape}
\usepackage{citesort}
\usepackage{amssymb}
\usepackage{appendix}
\usepackage{multirow}

\begin{document}
\def\qq{\langle \bar q q \rangle}
\def\uu{\langle \bar u u \rangle}
\def\dd{\langle \bar d d \rangle}
\def\sp{\langle \bar s s \rangle}
\def\GG{\langle g_s^2 G^2 \rangle}
\def\Tr{\mbox{Tr}}
\def\figt#1#2#3{
        \begin{figure}
        $\left. \right.$
        \vspace*{-2cm}
        \begin{center}
        \includegraphics[width=10cm]{#1}
        \end{center}
        \vspace*{-0.2cm}
        \caption{#3}
        \label{#2}
        \end{figure}
    }

\def\figb#1#2#3{
        \begin{figure}
        $\left. \right.$
        \vspace*{-1cm}
        \begin{center}
        \includegraphics[width=10cm]{#1}
        \end{center}
        \vspace*{-0.2cm}
        \caption{#3}
        \label{#2}
        \end{figure}
                }

\def\ds{\displaystyle}
\def\beq{\begin{equation}}
\def\eeq{\end{equation}}
\def\bea{\begin{eqnarray}}
\def\eea{\end{eqnarray}}
\def\beeq{\begin{eqnarray}}
\def\eeeq{\end{eqnarray}}
\def\ve{\vert}
\def\vel{\left|}
\def\ver{\right|}
\def\nnb{\nonumber}
\def\ga{\left(}
\def\dr{\right)}
\def\aga{\left\{}
\def\adr{\right\}}
\def\lla{\left<}
\def\rra{\right>}
\def\rar{\rightarrow}
\def\lrar{\leftrightarrow}
\def\nnb{\nonumber}
\def\la{\langle}
\def\ra{\rangle}
\def\ba{\begin{array}}
\def\ea{\end{array}}
\def\tr{\mbox{Tr}}
\def\ssp{{\Sigma^{*+}}}
\def\sso{{\Sigma^{*0}}}
\def\ssm{{\Sigma^{*-}}}
\def\xis0{{\Xi^{*0}}}
\def\xism{{\Xi^{*-}}}
\def\qs{\la \bar s s \ra}
\def\qu{\la \bar u u \ra}
\def\qd{\la \bar d d \ra}
\def\qq{\la \bar q q \ra}
\def\gGgG{\la g^2 G^2 \ra}
\def\q{\gamma_5 \not\!q}
\def\x{\gamma_5 \not\!x}
\def\g5{\gamma_5}
\def\sb{S_Q^{cf}}
\def\sd{S_d^{be}}
\def\su{S_u^{ad}}
\def\sbp{{S}_Q^{'cf}}
\def\sdp{{S}_d^{'be}}
\def\sup{{S}_u^{'ad}}
\def\ssp{{S}_s^{'??}}

\def\sig{\sigma_{\mu \nu} \gamma_5 p^\mu q^\nu}
\def\fo{f_0(\frac{s_0}{M^2})}
\def\ffi{f_1(\frac{s_0}{M^2})}
\def\fii{f_2(\frac{s_0}{M^2})}
\def\O{{\cal O}}
\def\sl{{\Sigma^0 \Lambda}}
\def\es{\!\!\! &=& \!\!\!}
\def\ap{\!\!\! &\approx& \!\!\!}
\def\md{\!\!\!\! &\mid& \!\!\!\!}
\def\ar{&+& \!\!\!}
\def\ek{&-& \!\!\!}
\def\kek{\!\!\!&-& \!\!\!}
\def\cp{&\times& \!\!\!}
\def\se{\!\!\! &\simeq& \!\!\!}
\def\eqv{&\equiv& \!\!\!}
\def\kpm{&\pm& \!\!\!}
\def\kmp{&\mp& \!\!\!}
\def\mcdot{\!\cdot\!}
\def\erar{&\rightarrow&}
\def\olra{\stackrel{\leftrightarrow}}
\def\ola{\stackrel{\leftarrow}}
\def\ora{\stackrel{\rightarrow}}
% .........................................................

\def\simlt{\stackrel{<}{{}_\sim}}
\def\simgt{\stackrel{>}{{}_\sim}}

% .........................................................

\title{
         {\Large
                 {\bf
                 }Properties of  $D_{s2}^*(2573)$ charmed-strange tensor meson
         }
      }

\author {\small K. Azizi$^{1,\dag}$ ,  H. Sundu$^{2,\ddag}$,  J.
Y. S\"{u}ng\"{u}$^{2,*}$, N. Yinelek$^{2,**}$
\\\small $^1$Department of Physics, Do\u gu\c s University, Ac{\i}badem-Kad{\i}k\"oy, 34722 Istanbul, Turkey\\
$^2$\small Physics Department, Kocaeli University, 41380 Izmit, Turkey\\
$^\dag$\small e-mail:kazizi@dogus.edu.tr \\
$^\ddag$\small e-mail:hayriye.sundu@kocaeli.edu.tr \\
$^*$\small e-mail:jyilmazkaya@kocaeli.edu.tr\\
$^{**}$\small e-mail:neseyinelek@gmail.com}
\date{}

\begin{titlepage}
\maketitle
\thispagestyle{empty}

\begin{abstract}
The mass and current coupling  constant of the $D_{s2}^* (2573)$ charmed-strange meson is calculated in the framework of two-point QCD sum rule approach.
Although the quantum numbers of this meson is not exactly known, its width and decay modes are consistent with $I(J^P)=0(2^+)$, which we consider to write the interpolating current used in our calculations.
Replacing the light strange quark with up or down quark we also compare the results with those of $D_{2}^*$ charmed tensor meson and estimate the 
order of SU(3) flavor symmetry violation.
\end{abstract}

~~~PACS number(s): 11.55.Hx, 14.40.Lb
\end{titlepage}

%%%
\section{Introduction}

During last few years  many new particles have been discovered in different experiments.
With increased running energies of colliders and improved
sensitivity of detectors, more hadrons are expected to be observed. To better understand and analyze the experimental results, parallel theoretical and phenomenological studies on the spectroscopy and decay properties of 
newly discovered particles are needed. The LHCb Collaboration at CERN reported first observation of the $D_{s2}^* (2573)$ particle through
 the semileptonic $\bar B_s^0\rightarrow D_{s2}^{*+}X\mu^-\bar \nu$ transition in 2011 \cite{LHCb}. This decay has an important contribution to the total branching ratio of the semileptonic $\bar B_s^0$  decays, 
so  its analysis helps us get more information about the semileptonic $\bar B_s^0$  decays which is less known experimentally compared to lighter $B$ mesons.

Although the quantum numbers of the observed $D_{s2}^* (2573)$ particle  is not exactly known, however, its width and decay modes favors the quantum number  $I(J^P)=0(2^+)$ \cite{pdg}. In this article, 
we calculate the mass and current coupling  constant of the $D_{s2}^* (2573)$ in the framework of two-point QCD sum rules considering it as a  charmed-strange tensor meson. The interpolating  currents
of the tensor mesons contain derivatives, so we calculate the two-point  correlation function first in coordinate space then transform calculations to the momentum space to apply Borel transformation and continuum subtraction
in order to isolate the ground state particle from the higher states and continuum. For some  experimental and theoretical works/reviews on the properties, structure and decay channels of charmed-strange mesons, see
for instance \cite{Fazio,Faessler,Bexperiment,Belleexp,colangelo,twostates,DsJ} and references therein.

The outline of the article is  as follows. Starting from an appreciate two-point correlation function, we derive QCD sum rules for the mass and current coupling constant of the  $D_{s2}^* (2573)$
charmed-strange tensor meson in next section. In section 3, we numerically analyze the sum rules obtained in section 2 and obtain working regions for auxiliary Borel parameter and continuum threshold entered to calculations. 
Making use of the working regions for auxiliary parameters, we obtain the numerical values of the mass and decay constant of the tensor meson under consideration. Replacing the strange quark with the up or down quark we also
find the masses and decay constant of the corresponding $\bar d(\bar u)c$ system, by comparison of which we estimate  the order of SU(3) flavor symmetry violation in the charmed tensor  system.

%%%
%%%
\section{Mass and current coupling  of $D_{s2}^*(2573)$ charmed-strange tensor meson}
Hadrons are formed in a range of energy very lower than the perturbative or asymptotic region, so to investigate their properties some non-perturbative approaches are required. Among
 non-perturbative methods  the QCD sum rule \cite{Shifman} is one of the most attractive and applicable tools to hadron physics as it is free of any model dependent parameters and is based on QCD Lagrangian. 
According to the philosophy of this model, to calculate the masses and current coupling constant, we start with a two-point correlation function and calculate it once in terms of hadronic parameters called the physical 
or phenomenological side, and the other in terms of QCD parameters in deep Euclidean region via operator product expansion (OPE) called the QCD or theoretical side. The QCD sum rules for the mass and current coupling constant are 
obtained matching both sides of the two-point correlation function under consideration. To stamp down the contribution of the higher states and continuum we apply Borel transformation to both sides of the acquired sum rules and use the quark-hadron duality
assumption.

To derive the QCD sum rules for physical quantities under consideration, we start with the following two-point
correlation function:
\begin{eqnarray}\label{correl.func.101}
\Pi _{\mu\nu,\alpha\beta}=i\int d^{4}xe^{iq(x-y)}{\langle}0\mid
{\cal T}[j _{\mu\nu}(x) \bar j_{\alpha\beta}(y)]\mid  0{\rangle},
\end{eqnarray}
where  ${\cal T}$ is the time-ordering
operator and  $j_{\mu\nu}$ is the interpolating current of the
$D_{s2}^*(2573)$ charmed-strange tensor meson. Considering the quantum numbers of $D_{s2}^*(2573)$ meson, its interpolating current in terms of quark fields  can be written as 
\begin{eqnarray}\label{tensorcurrent}
j _{\mu\nu}(x)=\frac{i}{2}\left[\bar s(x) \gamma_{\mu} \olra{\cal
D}_{\nu}(x) c(x)+\bar s(x) \gamma_{\nu}  \olra{\cal D}_{\mu}(x)
c(x)\right],
\end{eqnarray}
where the two-side covariant derivative $ \olra{\cal D}_{\mu}(x)$ is
defined as
\begin{eqnarray}\label{derivative}
\olra{\cal D}_{\mu}(x)=\frac{1}{2}\left[\ora{\cal D}_{\mu}(x)-
\ola{\cal D}_{\mu}(x)\right],
\end{eqnarray}
and
\begin{eqnarray}\label{derivative2}
\overrightarrow{{\cal
D}}_{\mu}(x)=\overrightarrow{\partial}_{\mu}(x)-i
\frac{g}{2}\lambda^aA^a_\mu(x),\nonumber\\
\overleftarrow{{\cal
D}}_{\mu}(x)=\overleftarrow{\partial}_{\mu}(x)+
i\frac{g}{2}\lambda^aA^a_\mu(x).
\end{eqnarray}
 Here $\lambda^a$   are the Gell-Mann matrices and $A^a_\mu(x)$ denote
the external  gluon fields. In  the Fock-Schwinger gauge, where
$x^\mu A^a_\mu(x)=0$, the external  gluon fields are expanded in terms of the
gluon field strength tensor as
\begin{eqnarray}\label{gluonfield}
A^{a}_{\mu}(x)=\int_{0}^{1}d\alpha \alpha x_{\beta}
G_{\beta\mu}^{a}(\alpha x)= \frac{1}{2}x_{\beta}
G_{\beta\mu}^{a}(0)+\frac{1}{3}x_\eta x_\beta {\cal D}_\eta
G_{\beta\mu}^{a}(0)+\cdots.
\end{eqnarray}
Note that we consider the currents in the aforementioned correlation function  at points $x$ and $y$, however, we have only  integral over four-$x$. 
The interpolating current of the tensor meson
 contains derivatives with respect to the space-time. 
Hence, after applying derivatives we will set $y=0$ then perform integral over  four-$x$.

In the physical side, the  correlation function in Eq.(\ref{correl.func.101}) is calculated by saturating it via a
complete set of states with the quantum numbers of  $D_{s2}^*(2573)$. After isolating the ground state and performing 
the four-integral we get
\begin{eqnarray}\label{phen1}
\Pi _{\mu\nu,\alpha\beta}=\frac{{\langle}0\mid j _{\mu\nu}(0) \mid
D_{s2}^*(2573)\rangle \langle D_{s2}^*(2573)\mid
\bar j_{\alpha\beta}(0)\mid
 0\rangle}{(m_{D_{s2}^*(2573)}^2-q^2)}
&+&\cdots,
\end{eqnarray}
where $\cdots$ symbolizes the contribution of higher states and
the continuum. To proceed,  we need  to define the matrix element
$\langle 0 \mid j_{\mu\nu}(0)\mid D_{s2}^*(2573)\rangle$ in terms of current coupling constant $f_{D_{s2}^*(2573)}$ and polarization tensor $\varepsilon_{\mu\nu}$:
\begin{eqnarray}\label{lep}
\langle 0 \mid j_{\mu\nu}(0)\mid
D_{s2}^*(2573)\rangle=f_{D_{s2}^*(2573)}
m_{D_{s2}^*(2573)}^3\varepsilon_{\mu\nu}.
\end{eqnarray}
Using    Eq.(\ref{lep}) in Eq.(\ref{phen1}) needs  performing
summation over polarization tensor, which is given as
\begin{eqnarray}\label{polarizationt1}
\varepsilon_{\mu\nu}\varepsilon_{\alpha\beta}^*=\frac{1}{2}\eta_{\mu\alpha}\eta_{\nu\beta}+
\frac{1}{2}\eta_{\mu\beta}\eta_{\nu\alpha}
-\frac{1}{3}\eta_{\mu\nu}\eta_{\alpha\beta},
\end{eqnarray}
where
\begin{eqnarray}\label{polarizationt2}
\eta_{\mu\nu}=-g_{\mu\nu}+\frac{q_\mu
q_\nu}{m_{D_{s2}^*(2573)}^2}.
\end{eqnarray}
As a result, for the the final expression of the physical side, we get
\begin{eqnarray}\label{pimunu}
\Pi_{\mu\nu,\alpha\beta}=\frac{f^2_{D_{s2}^*(2573)}m_{D_{s2}^*(2573)}^6}
{(m_{D_{s2}^*(2573)}^2-q^2)}
\left\{\frac{1}{2}(g_{\mu\alpha}~g_{\nu\beta}+g_{\mu\beta}~g_{\nu\alpha})\right\}+
\mbox{other structures}+\cdots
\end{eqnarray}
where we will choose  the explicitly written structure to extract the QCD sum rules for the mass and current coupling constant of the tensor meson.

In QCD side, the correlation function in Eq.(\ref{correl.func.101})  is calculated in the deep Euclidean
region where $q^2\ll0$, with the help of OPE where the short (perturbative)  and long
distance (non-perturbative)  contributions are separated. The perturbative part is
calculated using the perturbation theory, while the non-perturbative
part is parameterized in terms of QCD parameters such as quarks masses, quarks and gluon condensates, etc. Therefore,  any coefficient of the selected structure
 in QCD side
can be written as a dispersion integral plus a non-perturbative part:
\begin{eqnarray}\label{QCDside}
\Pi(q^2)=\int^{}_{}ds\frac{\rho^{pert}(s)}{(s-q^2)}+\Pi^{non-pert}(q^2),
\end{eqnarray}
where the spectral density  $\rho^{pert}(s)$ is obtained from the imaginary  of the perturbative contribution,  i.e.,  $\rho^{pert}(s)=\frac{1}{\pi}Im[\Pi^{pert}(s)]$. 

Our main goal in the following is to calculate the spectral density $\rho^{pert}(s)$ and the non-perturbative part $\Pi^{non-pert}(q^2)$.
Using the tensor current presented in
Eq.(\ref{tensorcurrent}) in the correlation function  in
Eq.(\ref{correl.func.101})  and contracting out all quark pairs
via the Wick's theorem, we obtain
\begin{eqnarray}\label{correl.func.2}
\Pi _{\mu\nu,\alpha\beta}&=&\frac{i}{4}\int d^{4}xe^{iq(x-y)}
\Bigg\{Tr\left[S_s(y-x)\gamma_\mu\olra{\cal D}_{\nu}(x)
\olra{\cal D}_{\beta}(y)S_c(x-y)\gamma_\alpha\right]\nonumber \\
&&+\left[\beta\leftrightarrow\alpha\right]
+\left[\nu\leftrightarrow\mu\right]+\left[\beta\leftrightarrow\alpha,
\nu\leftrightarrow\mu\right]\Bigg\}.
\end{eqnarray}
To proceed, we need to know
the expressions of the heavy and light quarks propagators, which are  calculated in \cite{L.J.Reinders}.
By ignoring from the gluon fields which have very small contributions to the mass and current coupling of the tensor meson (see also \cite{aliev,Kazim1,sundu}), the explicit
expressions of the heavy and light quarks propagators are given by 
\begin{eqnarray}\label{heavypropagator}
S_{c}^{ij}(x-y)=\frac{i}{(2\pi)^4}\int d^4k e^{-ik \cdot (x-y)}
\left\{ \frac{(\!\not\!{k}+m_c)}{(k^2-m_c^2)}\delta_{ij}
 +\cdots\right\} \, ,
\end{eqnarray}
and
\begin{eqnarray}
S_{s}^{ij}(x-y)&=& i\frac{(\!\not\!{x}-\!\not\!{y})}{
2\pi^2(x-y)^4}\delta_{ij}
-\frac{m_s}{4\pi^2(x-y)^2}\delta_{ij}-\frac{\langle
\bar{s}s\rangle}{12}\Big[1 -i\frac{m_s}{4}
(\!\not\!{x}-\!\not\!{y})\Big]\delta_{ij}
\nonumber\\
&-&\frac{(x-y)^2}{192}m_0^2\langle
\bar{s}s\rangle\Big[1-i\frac{m_s}{6}
(\!\not\!{x}-\!\not\!{y})\Big]\delta_{ij} +\cdots \, .
\end{eqnarray}

The next step is to use the expressions of the quarks propagators and
apply  derivatives with respect to $x$ and $y$ in
Eq.(\ref{correl.func.2}).  As a result, after setting $y=0$,  for the
 QCD side of the correlation function
in coordinate space, we get 
\begin{eqnarray}\label{correl.func.3}
\Pi_{\mu\nu,\alpha\beta}&=&\frac{~N_c}{16}
 \int\frac{d^4k}{(2\pi)^4}\frac{1}{(k^2-m_c^2)}\int
d^{4}xe^{i(q-k)\cdot
x}\Bigg\{\left[Tr\Gamma_{\mu\nu,\alpha\beta}\right]+
\left[\beta\leftrightarrow\alpha\right]\nonumber\\&+&\left[\nu\leftrightarrow\mu\right]+
\left[\beta\leftrightarrow\alpha,\nu\leftrightarrow\mu\right]\Bigg\},\nonumber\\
\end{eqnarray}
where $N_c=3$ is the color factor and the function $\Gamma_{\mu\nu,\alpha\beta}$ is given by 
\begin{eqnarray}\label{fonk}
&&\Gamma_{\mu\nu,\alpha\beta}\nonumber\\&=&k_{\nu}k_{\beta}\Big[\frac{i\!\not\!{x}}{2\pi^2x^4}+\frac{m_s}{4\pi^2x^2}+\Big(\frac{1}{12}+\frac{im_s\!\not\!{x}}{48}
+\frac{x^2m_0^2}{192}+\frac{ix^2m_sm_0^2\!\not\!{x}}{1152}\Big)\langle\bar{s}s\rangle\Big]\gamma_{\mu}(\!\not\!{k}+m_c)\gamma_{\alpha}\nonumber\\
&+&ik_{\beta}\Big[\frac{i}{2\pi^2}\Big(\frac{4x_{\nu}\!\not\!{x}}{x^6}-\frac{\gamma_{\nu}}{x^4}\Big)+\frac{m_sx_\nu}{2\pi^2x^4}
+\Big(\frac{im_s\gamma_\nu}{48}-\frac{x_\nu~m_0^2}{96}-\frac{im_0^2m_s(2x_\nu\!\not\!{x}+x^2\gamma_\nu)}{1152}
\Big)\langle\bar{s}s\rangle\Big]\nonumber\\
&\times&\gamma_\mu(\!\not\!{k}+m_c)\gamma_{\alpha}+\Big[\frac{4i}{2\pi^2x^6}\Big(\frac{6x_{\beta}x_{\nu}\!\not\!{x}}{x^{2}}+g_{\nu\beta}\!\not\!{x}-\gamma_{\nu}x_{\beta}
+\gamma_{\beta}x_{\nu}\Big)+\frac{m_s}{2\pi^2}\Big(-\frac{4x_{\beta}x_{\nu}}{x^6}+\frac{g_{\nu\beta}}{x^4}\Big)\nonumber\\
&+&\Big(\frac{g_{\nu\beta}m_0^2}{96}+\frac{im_sm_0^2}{576}\Big(g_{\nu\beta}\!\not\!{x}+x_{\beta}\gamma_{\nu}+x_{\nu}\gamma_{\beta}\Big)\Big)
\langle\bar{s}s\rangle\Big]\gamma_{\mu}(\!\not\!{k}+m_c)\gamma_{\alpha}-ik_\nu\Big[\frac{i}{2\pi^2}
\Big(\frac{\gamma_\beta}{x^4}-\frac{4x_\beta\!\not\!{x}}{x^6}\Big)\nonumber\\
&-&\frac{m_sx_\beta}{2\pi^2x^4}+\Big(\frac{im_s\gamma_\beta}{48}+\frac{m_0^2x_\beta}{96}+\frac{im_sm_0^2}{1152}
\Big(2x_\beta\!\not\!{x}+x^2\gamma_\beta\Big)\langle\bar{s}s\rangle\Big)\Big]\gamma_\mu(\!\not\!{k}+m_c)\gamma_{\alpha}\nonumber\\
&+&\left[\beta\leftrightarrow\alpha\right]+\left[\nu\leftrightarrow\mu\right]+\left[\beta\leftrightarrow\alpha,\nu\leftrightarrow\mu\right].
\end{eqnarray}
After performing all traces in Eq.(\ref{correl.func.3}), in order
to calculate the integrals, first we transform the
terms containing $\frac{1}{(x^2)^n}$  to the momentum space 
and replace $x_{\mu}\rightarrow -i\frac{\partial}{\partial
q_{\mu}}$. The integral over four-$x$ gives us a Dirac Delta
function, making use of which  we perform the integral over
four-$k$. To perform the final integral  over four-$p$ we use
the Feynman parametrization method and the relation
\begin{eqnarray}\label{Int}
\int d^4p\frac{(p^2)^{\beta}}{(p^2+L)^{\alpha}}=\frac{i \pi^2
(-1)^{\beta-\alpha}\Gamma(\beta+2)\Gamma(\alpha-\beta-2)}{\Gamma(2)
\Gamma(\alpha)[-L]^{\alpha-\beta-2}}.
\end{eqnarray}
After lengthy calculations for  the spectral density, we get
\begin{eqnarray}\label{Rhopert}
\rho^{pert}(s)&=&N_c\frac{(m_{c}^2-s)^3(2m_{c}^4+m_{c}^2s+10m_{c}m_{s}s-3s^2)}{960\pi^2s^3}.
\nonumber\\
\end{eqnarray}
For the non-perturbative part, we also obtain
\begin{eqnarray}\label{Rhononpert}
\Pi^{non-pert}(q^2)=m_0^2\langle
\bar{s}s\rangle\frac{(24m_c^3-m_c^2m_s-24m_cq^2-5m_sq^2)}{1152(m_c^2-q^2)^2}.
\end{eqnarray}

After acquiring the correlation function in both phenomenological and QCD sides, by the procedures mentioned in the beginning of this section, we obtain the following sum rule for the mass and current coupling of
the  $D_{s2}^*(2573)$ tensor meson:
\begin{eqnarray}\label{rhomatching}
f^2_{D_{s2}^*(2573)} e^{-m_{D_{s2}^*(2573)}^2/M^2}
=\frac{2}{m_{D_{s2}^*(2573)}^6}\Big(\int_{(m_c+m_s)^2}^{s_0} ds
\rho(s)~e^{-s/M^2}+\mathbf{\hat{B}}\Pi^{non-pert}(M^2)\Big),
\end{eqnarray}
where $s_0$ is the continuum threshold and $M^2$ is the Borel mass
parameter. The function $\mathbf{\hat{B}}\Pi^{non-pert}(M^2)$ in Borel scheme is obtained as
\begin{eqnarray}
\mathbf{\hat{B}}\Pi^{non-pert}(M^2)=m_0^2\langle\overline{s}s
\rangle\frac{(24M^2m_c+5M^2m_s-6m_c^2m_s)}{1152M^2}e^{-m_c^2/M^2}.
\end{eqnarray}
The mass of the $D_{s2}^*(2573)$ tensor meson alone is obtained from 
\begin{eqnarray}\label{mass}
m_{D_2^*(2573)}^2 =\frac{\int_{(m_c+m_s)^2}^{s_0}ds~s~\rho(s)e^{-s/M^2}+\frac{\partial}{\partial(-1/M^2)}\mathbf{\hat{B}}\Pi^{non-pert}(M^2)}
{\int_{(m_c+m_s)^2}^{s_0}ds\rho(s)~
e^{-s/M^2}+\mathbf{\hat{B}}\Pi^{non-pert}(M^2)}.
\end{eqnarray}
\section{Numerical results}
In this section, we numerically analyze the sum rules obtained for the mass and current coupling constant of the $D_{s2}^*(2573)$ tensor meson in the previous section. For this aim we use some input
parameters as $m_c=(1.275\pm 0.025)~GeV$
\cite{pdg}, $\langle \bar
ss(1~GeV)\rangle=-0.8(0.24\pm0.01)^3~GeV^3$ \cite{B.L.Ioffe} and
$m_0^2(1~GeV) = (0.8\pm0.2)~GeV^2$ \cite{m02}. 

The
 sum rules for above mentioned physical quantities also contain two auxiliary parameters: the Borel
 parameter $M^2$ and the continuum threshold $s_0$ coming from the Borel transformation and the continuum subtraction, respectively. In the following, we shall find working regions of these parameters such that the 
results of the mass and current coupling show weak dependences on these auxiliary parameters according to the general criteria of the method.  The continuum threshold
$s_{0}$ is not completely capricious, but it is correlated with the
energy of the first excited state with the same quantum numbers. As a result, we choose 
$s_0=(10.0\pm0.5)~GeV^2$ for the continuum threshold.

\begin{figure}[h!]
\begin{center}
\includegraphics[totalheight=8cm,width=10cm]{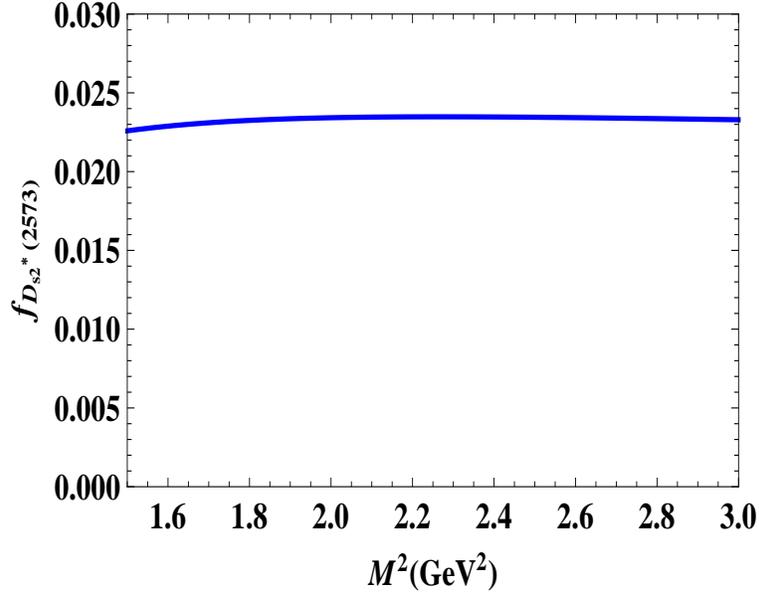}
\end{center}
\caption{The dependence of current coupling $f_{D_{s2}^*(2573)}$ on
Borel mass parameter $M^{2}$ at $s_0=10~GeV^2$.} \label{f1Msqgraph}
\end{figure}
\begin{figure}[h!]
\begin{center}
\includegraphics[totalheight=8cm,width=10cm]{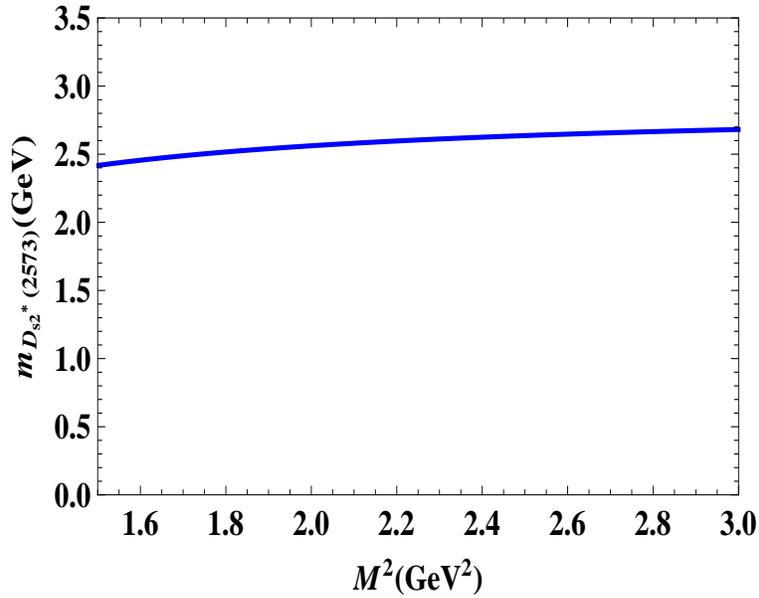}
\end{center}
\caption{The dependence of  mass of  $D_{s2}^*(2573)$ on
Borel mass parameter $M^{2}$ at $s_0=10~GeV^2$.} \label{f1Msqgraph}
\end{figure}

 The working
region for the Borel mass parameter is calculated  demanding that not only the contributions of the  higher state and continuum
are stamped down but also the contribution of the
higher order operators are very small. The later means that the series of sum rules for physical quantities are convergent and the 
perturbative part constitutes an important part of the whole contribution. In other words, the upper bound on the Borel  parameter is found by demanding that 
\bea
\label{nolabel}
{\ds \int_{s_{min}}^{s_0}\ds  \rho(s) e^{-s/M^2} \over
\ds \int_{s_{min}}^\infty \rho(s) e^{-s/M^2}} ~~>~~ 1/2, 
\eea
 which leads to

\bea
\label{e8202}
M_{max}^2 =
3~GeV^2.
\eea

The lower bound on this parameter is obtained requiring that the contribution of the 
perturbative part exceeds   the nonperturbative
contributions. From this restriction we get  
\bea
\label{e8202}
M_{min}^2 =
1.5~GeV^2.
\eea
We depict the dependence of the current coupling constant and mass of the tensor meson under consideration on Borel mass parameter at a fixed value of continuum threshold in figures 1 and 2. From these figures
 we see that the results weakly depend on the 
Borel mass parameter in its working region. Here, we would like to make the following comment. The above analyses have been done based on, so
called, the standard procedure in QCD sum rule technique such that  the quark-hadron duality assumption as a systematic error has been used and the continuum threshold has been taken independent of Borel mass parameter. However, as also stated in \cite{melikov}, the continuum threshold 
can depend on $M^2$. Hence the standard procedure   does not render the  realistic errors and, in fact, the actual errors should be large. 
Our numerical calculations show that taking the continuum threshold dependent on Borel mass parameter brings an extra systematic error of $\%15$, which we will add to our numerical values.
 
Making use of the working regions for auxiliary parameters and taking into account all systematic uncertainties,  we obtain the 
numerical results of the mass and current coupling constant  for
$D_{s2}^*(2573)$ tensor meson as presented  in Table \ref{tab1}. We also compare our result on the mass with the existing experimental data which shows a good consistency.
The errors quoted in our predictions belong to the uncertainties in determination of  working regions for 
both auxiliary parameters, those existing in other inputs as well as systematic errors. Our result on the current coupling constant of the charmed strange $D_{s2}^*(2573)$ tensor meson can be checked in future experiments.
\begin{table}[h] \centering
 \begin{tabular}{|c||c|c|c|c|} \hline &
 Present Work&Experiment \cite{pdg}
\\\cline{1-3}\hline\hline
$m_{D_{s2}^*(2573)}$& $(2549\pm440)~MeV $ &$(2571.9\pm0.8)~MeV$\\
\cline{1-3} $f_{D_{s2}^*(2573)}$& $0.023\pm0.011~$ &-\\
\cline{1-3}\hline\hline
\end{tabular}
\vspace{0.8cm} \caption{Values for the mass and current coupling constant of
the $D_{s2}^*(2573)$ tensor meson.} \label{tab1}
\end{table}
%
%The errors quoted in our predictions are due to the variations of
%both auxiliary parameters and uncertainties in input parameters.
%From Table \ref{tab1}, we see a good consistency between our
%prediction and  the experimental data on the masses of the
%$D_{s2}^*(2573)$ tensor meson.
% In addition, we draw the decay
%constant versus $M^2$ graph in Figure 1. As shown in the Figure,
%the most important contribution to the decay constant comes from
%the perturbative part.

Our final goal is to replace the strange quark with the up/down quark and estimate the order of  SU(3) flavor violation. Our calculations show that this violation is maximally $\%7$ in 
 the case of charmed tensor meson.

%We also compare our numerical results for the decay constant with
%the ref. \cite{HayKaz} and see that $D_{s2}^*(2573)$ meson decay
%constant is lower $\%20$ than the $D_{s2}^*(2460)$ state.

%The analysis of the $D_{s2}^*(2573)$ heavy-light meson is an
%important source for the determination of its quantum numbers.
%Precise determination of $D_{s2}^*(2573)$ meson parameters can be
%used to classify observed states in heavy quark doublets. The
%confirmation of our predictions is expected in the very near
%future from the experiments. Extensive studies of $D_{s2}^*(2573)$
%tensor meson have been carried out at LHC and may contain
%promising data.

\section{Acknowledgment}
This work has been supported in part by the Kocaeli University
fund under BAP project No. 2011/119.

\end{document}